\documentclass{sig-alternate-05-2015}
\usepackage{etoolbox}
\makeatletter
\patchcmd{\maketitle}{\@copyrightspace}{}{}{}
\makeatother
\usepackage{graphicx}
\usepackage{epstopdf} 
\usepackage{subcaption}
\usepackage[labelfont=bf]{caption}
\usepackage{ntheorem}
\newtheorem{hyp}{Hypothesis}
\usepackage[inline]{enumitem}

\begin{document}






%

\title{Perceived Performance of Webpages In the Wild}

\subtitle{\vspace{-0.2in}Insights from Large-scale Crowdsourcing of Above-the-Fold QoE}

%
%
%
%
%

\numberofauthors{3} 
%
\author{
\alignauthor \vspace{-0.2in}Qingzhu Gao*\\    
\alignauthor \vspace{-0.2in}Prasenjit Dey\\      
\alignauthor \vspace{-0.2in}Parvez Ahammad*\\     
\and 
      \affaddr{Instart Logic Inc., 450 Lambert Ave, Palo Alto, CA, USA}  \\
      \email{\{qgao, pdey, pahammad\}@instartlogic.com; * = Equal contribution}
}
\maketitle
\begin{abstract}
Clearly, no one likes webpages with poor quality of experience (QoE). Being perceived as slow or fast is a key element in the overall perceived QoE of web applications. While extensive effort has been put into optimizing web applications (both in industry and academia), not a lot of work exists in characterizing what aspects of webpage loading process truly influence human end-user's perception of the \emph{Speed} of a page. In this paper we present \emph{SpeedPerception}, a large-scale web performance crowdsourcing framework focused on understanding the perceived loading performance of above-the-fold (ATF) webpage content. Our end goal is to create free open-source benchmarking datasets to advance the systematic analysis of how humans perceive webpage loading process. 

In Phase-1 of our \emph{SpeedPerception}  study using Internet Retailer Top 500 (IR 500) websites~\cite{github}, we found that commonly used navigation metrics such as \emph{onLoad} and \emph{Time To First Byte (TTFB)} fail (less than 60\% match) to represent majority human perception when comparing the speed of two webpages. We present a simple 3-variable-based machine learning model that explains the majority end-user choices better (with $87 \pm 2\%$ accuracy). In addition, our results suggest that the time needed by end-users to evaluate relative perceived speed of webpage is far less than the time of its \emph{visualComplete} event. 
\end{abstract}

%
%


%
%

%
%



\section{Introduction}
Bad quality of experience (QoE) is not just annoying to end-users, but also costly for website owners. A recent survey indicated that 49\% of users will abandon a site after experiencing performance issues and that a 1-second delay meant a 7\% reduction in conversions~\cite{6smarketing}. 
Page-level navigation metrics (e.g. onLoad, TTFB) are typically thought to not only reflect the speed of application-level delivery pipeline, but also have direct impact on the business for E-Commerce websites \cite{catchpoint}. 

Improving \emph{onLoad} (or other performance metrics) became a popular area of research in recent years~\cite{iyengar2005web, cohen2002prefetching, ahammad2015flexible, netravali2016polaris, wang2016speeding}. 
Unfortunately, none of these techniques directly take real end-user experience into account. 
\cite{butkiewicz2011understanding,balachandran2014modeling} studied the underlying pattern of web page complexity on user experience. Recently, \cite{kelton17improving} took an important step towards estimating user's perception using eyeball-tracking technology. Lack of ability to account for dynamic third-party contents and being specific to one website at a time are the main drawbacks for \cite{kelton17improving}. 
Google has put forth SpeedIndex~\cite{speedIndex} to replace traditional W3C metrics for measuring \emph{above-the-fold} content performance. 
\cite{bocchi2016measuring} introduced two SpeedIndex like metrics and described the correlation among them. They assumed that SpeedIndex alone is sufficient to account for end-user QoE without any end-user validation.

\cite{varvello2016eyeorg} created an experiment that allowed users to look at the webpage loading frame-by-frame to determine the user perceived page load time (UPPLT). However, understanding the relationship between static measurements and user experience is non-trivial and hard to generalize across websites for a key reason: a user's perception of speed (when presented a single webpage in isolation) is subjective \cite{egger2012time}. For example, consumers may tolerate a local small business site loading in 5 seconds, but they may not wait the same amount of time if they were browsing a top-tier popular webpage, since their expectations are different.  
We want to fill this critical gap by creating an A/B comparison framework and identify the most relevant metric(s) that explain user perception across a broad swath of commercial websites. 

We built
\emph{SpeedPerception}~\footnote{http://www.speedperception.com} to crowdsource users' perceived performance of AFT webpage content. Our aim is to
enable reproducible research that improves understanding of the web application QoE at scale. The advantage of our comparative paradigm is that we can resolve the scalability limitations of small grouped experiments~\cite{hossfeld2014best}. 
Our belief (and hope) is that \emph{SpeedPerception}-like benchmarking datasets can provide a quantitative basis to compare different algorithms and spur progress on helping quantify perceived webpage performance. 




\section{Metrics for Webpage QoE}
In the past years, the web performance community settled on Page Load Time (\emph{onLoad}) and a few other page-level navigation metrics for evaluating the performance (QoE) of a webpage. An often-repeated industrial dogma is that, given the same network conditions, content structure, and other controllable factors, the smaller these metrics, the better the QoE of a webpage (from an end-user perspective). Some recent studies have proposed new set of metrics (such as byteIndex~\cite{bocchi2016measuring}, and UPPLT~\cite{varvello2016eyeorg}) for measuring end-user QoE. 
While new metrics continue to be proposed, what is sorely missing is a systematic and reproducible way to link them 
via real human feedback. \emph{SpeedPerception} fills this gap. Using \emph{SpeedPerception}, we have attempted to quantify how accurate the current web performance metrics are in representing real user judgments of perceived speed. 

The metrics included in the \emph{SpeedPerception} study are mostly defined by WebPagetest \footnote{https://www.webpagetest.org/} and W3C \footnote{https://www.w3.org/TR/navigation-timing/}, plus one novel metric, Perceptual SpeedIndex (PSI).
We also explore novel variations to SI and PSI through changing the end point of their integrals in a systematic way. We group these synthetic metrics into two categories: \emph{non-visual} and \emph{visual} metrics. 

\subsubsection*{Non-Visual Metrics} 
\textbf{\emph{Time to First Byte (TTFB)}} is  the time from the initial navigation until the first byte is received by the browser. \textbf{\emph{DOM Content Load Event End (DCLend)}} is time at which the DOM has been loaded by parsing the response. \textbf{\emph{onLoad (Load Time or PLT)}} is measured as the time from the start of the initial navigation until the beginning of the window load event.

\subsubsection*{Visual Metrics}
\textbf{\emph{First Paint}} is a measurement reported by the browser itself about when it thinks it painted the first content. It is available from JavaScript and can be reported from the field.
\textbf{\emph{Render Start (render)}} is the time from the start of the initial navigation until the first non-white content is painted. It is measured by capturing video of the page load and looking at each frame for the first time the browser displays something other than a blank page. It can only be measured in a lab and is generally the most accurate measurement for it. \textbf{\emph{SpeedIndex (SI)}} is the average time at which visible parts of the page are displayed. It measures how quickly the page contents are visually populated (where lower numbers are better). \textbf{\emph{Visual Complete (visualComplete)}} is the time from the start of the initial navigation until there is no visual process within above-the-fold content.


\subsubsection*{Perceptual SpeedIndex (PSI)}
Google introduced the SI in 2012 as a metric to measure above-the-fold (ATF) visual QoE. The main idea was to use an aggregate function on the quickness of ATF visual completion process. The frame-to-frame visual progress in SI is computed from pixel-histogram comparisons. SI's histogram-based visual progress calculations can also lead to some issues. In particular, visual jitter (caused by layout instability, weird ad behavior or carousel elements, etc.) cannot be captured by pixel-histogram based calculations. 

To address these issues, we proposed PSI as a complementary visual QoE metric to serve as a proxy for end-user perception\footnote{http://www.parvez-ahammad.org/blog/perceptual-speed-index-psi-for-measuring-above-fold-visual-performance-of-webpages}. Empirical experiments demonstrated that PSI and SI are linearly correlated with a strong correlation score of 0.91. Despite the strong correlation, SI and PSI are actually complementary to each other. While SI focuses on addressing how most of the webpage ATF content loads quickly, PSI focuses on addressing if most of the webpage ATF content loads quickly without visually noticeable jitter. Since October 2016~\footnote{https://github.com/GoogleChrome/lighthouse/pull/785}, Google Chrome has officially incorporated PSI in their LightHouse project for measuring Progressive Web App performance~\cite{lighthouse}.

The math behind SI and PSI is relatively simple: aggregating the visual progress along a web page loading time-line. This can be expressed as:
\begin{equation}
    \label{SIEquation}
    {Index}_{end} = \int_{start}^{end} 1 - \frac{\text{Visual Completeness}}{100} dt
\end{equation}
\emph{Visual Completeness} in \textbf{Eq \ref{SIEquation}} is with respect to the last frame of a webpage loading process (video-based), and is expected to be $0$ at $t=\text{start}$, and $100$ at $t=\text{end}$. Instead of using mean pixel-histogram difference (MHD) to measure visual completeness, PSI uses Structural Similarity \cite{wang2004image}. Both SI and PSI start the integration at time 0 and end at visualComplete. In practice, truncating this integral at the right endpoint can make a big difference. For example, $SI_{onLoad}$ would be defined as:
\begin{equation}
    \label{SIonLoadEquation}
    {SI}_{onLoad} = \int_{start}^{onLoad} 1 - \frac{\text{Visual Completeness (MHD)}}{100} dt
\end{equation}
$SI_{TTC}$ and $PSI_{TTC}$ can similarly be defined with respect to Time to Click (TTC, Section 5.1): 
\begin{equation}
    \label{PSIttcEquation}
    {PSI}_{TTC} = \int_{start}^{TTC} 1 - \frac{\text{Visual Completeness ({SSIM})}}{100} dt
\end{equation}

\section{SpeedPerception: Phase-1}
\emph{SpeedPerception} is built to serve the purpose of our crowdsourcing experiments where ATF loading process of two webpages are displayed to participants and their responses are recorded. 
The \emph{SpeedPerception} framework 
enables the study of how human end-users perceive the ``faster'' page given two choices. 
It is important to conduct such experiments via large-scale crowdsourcing such that (1) there is sufficient sample size on each comparison; (2) there is coverage across a large population of participants so that we have distinct subjective opinions.

Consider the question: \emph{How does one evaluate end-user perception of webpage speed?} As subjective as this sounds, perception of webpage speed is also relative (A vs B). Participants in our study were expected to answer a simple question, \emph{which web page do you perceive to be faster?}
To ensure the identical visual experience, we used videos that are generated from WebPagetest. These videos capture the above-the-fold content rendering process. 

The web application is built with the Meteor.Js framework.
We use MongoDB as the backend database. Both the application and database are hosted on (separate) cloud servers that can be scaled up or down dynamically. We have open-sourced our entire framework on Github \footnote{https://github.com/pdey/SpeedPerceptionApp} with the  MIT license to facilitate reproducibility and ease of use.

\subsection{Hypotheses and Workflow}
Three testable experimental hypotheses were set up before the Phase-1 experiment:
\begin{hyp}
\textbf{No single existing performance metric can explain an end-user's perception of Speed with above 90\% accuracy.}
\end{hyp}
\begin{hyp}
\textbf{Visual metrics will perform better than non-visual metrics in predicting end-user's judgments on ATF performance.}
\end{hyp}
\begin{hyp}
\textbf{An end-user will not wait until \emph{visualComplete} event to make their choice.}
\end{hyp}

\begin{figure}
\graphicspath{ {/} }
\centering
\includegraphics[scale=0.4]{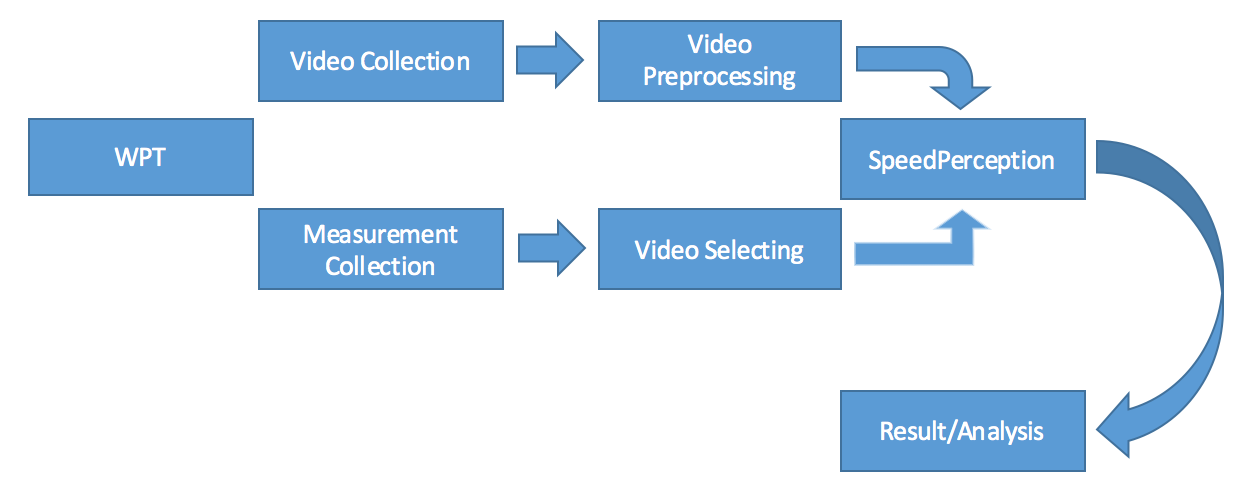}
\vspace{-0.2in}
\caption{\textbf{SpeedPerception workflow, from data collection to analysis.}}
\label{fig:workflow}
\vspace{-0.2in}
\end{figure}

\begin{figure*}
\graphicspath{ {/} }
\begin{subfigure}{.6\textwidth}
  \centering
  \includegraphics[scale=0.25]{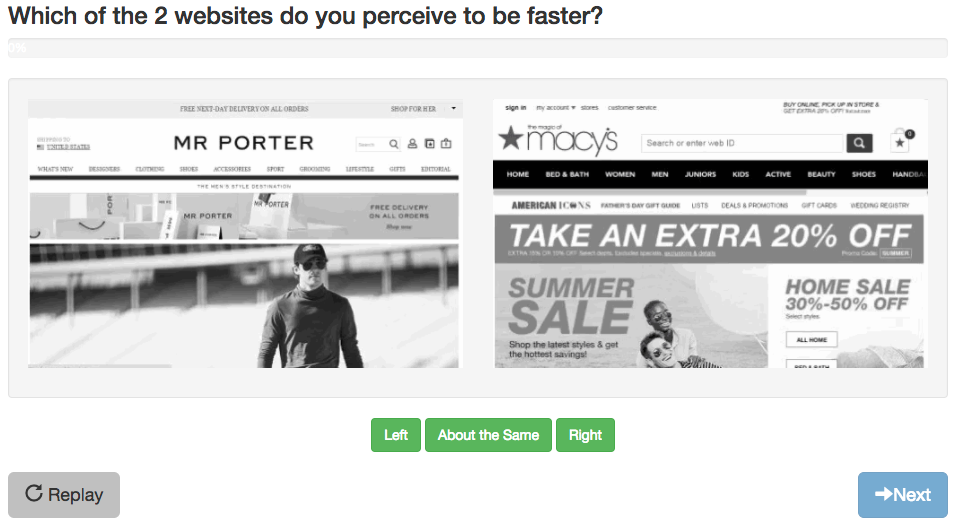}
  \caption{\textbf{Side-by-side layout for each pair of videos.}}
  \label{fig:UI}
\end{subfigure}%
\begin{subfigure}{.4\textwidth}
  \centering
  \includegraphics[scale=0.25]{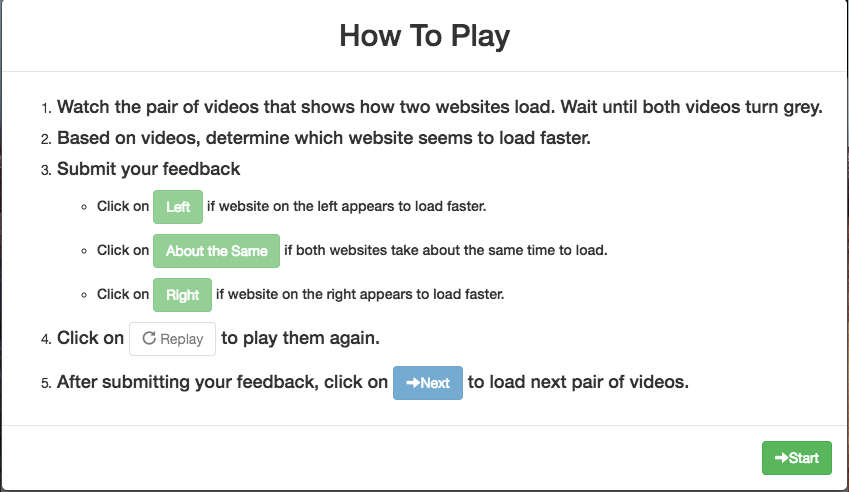}
  \caption{\textbf{Instructions for participation.}}
  \label{fig:Instruction}
\end{subfigure}
\vspace{-0.1in}
\caption{\textbf{SpeedPerception user interface}}
\vspace{-0.2in}
\label{fig:userInterface}
\end{figure*}

Figure \ref{fig:workflow} shows the work flow of SpeedPerception. First,  we collect static measurements (HAR or \emph{HTTP Archive} files) and videos from WebPagetest. 
Videos and HARs need to be processed so that we have a more structured data. Let us first discuss some key characteristics of Video and HAR files. 

\textbf{\emph{HAR}} On an private instance (e.g., EC2 on Amazon) of WebPagetest, we launched a series of tests on Internet Retailer Top 500 (IR 500) URLs \footnote{https://www.digitalcommerce360.com/product/top-500/}. Test configurations are consistent across all test runs; we used ``Chrome 50.0.2661.102,'' ``Cable 10/5 Mbps,'' and ``North California'' machines. We ran 10 unique tests on each URL to have a fair sampling and reduce outliers caused by network hang-ups, traffic spikes and other factors. HAR file returned from each test contains the log of a web browser's interaction with the site when loading it. 


\textbf{\emph{Video}} A video of the ATF content is associated with each HAR file so that we can later map performance metrics to end-users' perception of ``Speed.'' WebPagetest makes it possible to record a live webpage video while running a test. The cut-off was set at \emph{visualComplete} of these videos. Most videos have length (time) less than the actual ``Fully Loaded' simply because there is more content to be loaded below users' view-ports.

\subsection{Video Pair Selection}
We then applied a group of 16 conditions to generate our target videos. In order to fairly compare between pairs, we need to pay attention to ``Visual Complete.'' Controlling for the end point of a video, we only selected video pairs within 5\% normalized difference of \emph{visualComplete} (vc). The normalized difference is calculated as: 
\begin{equation}
    \label{diff_equation}
    \text{diff}(vc_1, vc_2) = \frac{(vc_1 - vc_2)}{(vc_1 + vc_2)*0.5}
\end{equation}


Within 5\% \emph{visualComplete} difference, we subgroup them based on 4 conditions of SI difference: 
$ SI_{diff} >= 10 $;
$ 1 <= SI_{diff} < 10 $;
$ -10 < SI_{diff} <=- 1 $;
$ SI_{diff} <=- 10 $.

Within each SI difference condition, we again subgroup each of them into 4 conditions of PSI difference: 
$ PSI_{diff} >= 10 $;
$ 1 <= PSI_{diff} < 10 $;
$ -10 < PSI_{diff} <=- 1 $;
$ PSI_{diff} <=- 10 $.

In total, we have $4 * 4 = 16$ conditions for our experiment. The reason we selected our video pairs based on SI and PSI for Phase-1 experiment is that we believed these are key QoE metrics to best express user perception. We also selected 5 fixed ``honeypot pairs'' (see section 4) in addition to 10 sets of 16 video pairs (total of 160 pairs + 5 honeypots). The final 160 pairs came from 115 unique webpages. 

\subsection{Platform Design}

\textbf{Figure \ref{fig:UI}} shows the UI of SpeedPerception. A typical session begins with an instruction banner (\textbf{Figure \ref{fig:Instruction}}), which participants are expected to carefully read and follow these steps during the experiment. After clicking the \textbf{Start} button, a total number of 21 pairs of videos will be displayed sequentially, 16 assessment pairs + 5 honeypot pairs. For each instance, 2 videos start at the same time and play in parallel side by side. The parallel layout allows participants to better evaluate the webpage loading process. After watching a video pair, we provide 3 options for users to report their response. They can pick ``Left'' or ``Right'' if they perceived one of the webpages to load faster, otherwise they can pick ``About the same'' when unable to determine a ``winning'' candidate. The \textbf{Replay} button enables participants to replay the video pairs as many times as they want until they feel comfortable to make a choice. After reviewing the pair, user will click on \textbf{Next} to proceed. 

We randomly chose a set of video pairs out of 10 to present to our participants. 
We assigned a unique session ID to every single attempt from participants who clicked on the ``Start'' button, because one could have perceived each session differently given the randomized selection of video pairs.



\section{Engagement Validation}
One of the key challenges of crowdsourcing is to ensure the quality of the data. \emph{SpeedPerception} experiment did not acquire participants from any known platform such as Amazon Mechanical Turk or Microworkers because we simply do not have faith in paying people to provide webpage QoE judgments. We promoted our experiment mostly through social media channels in the web performance community, as well as colleagues and friends. To mitigate contamination of the data, or malicious responses, we set up a series of validation mechanisms. 

\textbf{\emph{Instructions}}: Participants are given clear instructions as shown in \textbf{Figure \ref{fig:Instruction}}. We expected people to follow these rules, except we did accept decisions/feedback before both videos reach \emph{visualComplete}.

\textbf{\emph{Enforcement}}: During any stage of the experiment, participants cannot skip to the next video pair without providing a choice first. We want every video pair to be assessed --- so the ``Click'' button was not made available until a choice had been made. Failure to complete all 21 pairs in a given session was considered as invalid session. Data points from invalid sessions were excluded from our analysis.


\textbf{\emph{Honeypots}}: To prevent any malicious participant or bot, we used a ``honeypot'' mechanism. We inserted 5 video pairs at random order with known (very obvious) choices in each session of the study. One honeypot mistake is allowed per session, so that we only take responses that exceed 80\% or more on these honeypots.  

\textbf{\emph{Majority Vote}}: For each of 160 (excluding 5 honeypots) video pairs, we aggregate across the participants' votes to formulate a ``majority vote''. 
For example, if a given pair has 10\% votes for ``Left'', 30\% votes for ``Equal'' and 60\% votes for ``Right'', then we consider ``Right'' video as the majority human choice.

\section{Results \& Analysis}
A total number of 5,400+ sessions were recorded in the \emph{SpeedPerception} Phase-1 experiment, during a period of 2 months. 51\% of the sessions successfully finished all 21 evaluations and passed the ``honeypot' threshold. Accordingly, we have more than 40,000 valid votes that are nicely distributed over the 160 video pairs, with 250+ votes on each pair. Each video pair has votes split between 3 choices.  
47\% of the majority votes are on ``Left'' and 46\% on ``Right'', while 7\% of them fall in ``Equal''. Participants seem to have a strong preference to pick one of the two sides, instead of  ``Equal'', when comparing between two webpages. It also indicates that our video pairs were fairly selected and displayed without any bias to one side. 

Our primary goal is to examine the pattern of how page-level performance metrics 
reflect user perception. We calculated the normalized difference on each metric for every pair, using Equation \ref{diff_equation}. Then a ``synthetic vote'' was assigned for each video pair using the normalized difference of these metrics, where difference falls into $\pm 5\%$ will be assigned to ``Equal.'' Difference smaller than -5\% will considered as ``Left,'' and larger than +5\% as ``Right.'' We also tried different thresholds using $\pm 10\%$ and $\pm 1\%$, just to make sure that the choice of threshold doesn't have an undue impact on the results. We found that the overall results do not change significantly. For conciseness, we only include plots using $\pm 5\%$ in this paper.

\begin{figure}
\graphicspath{ {/} }
\centering
\includegraphics[scale=0.3]{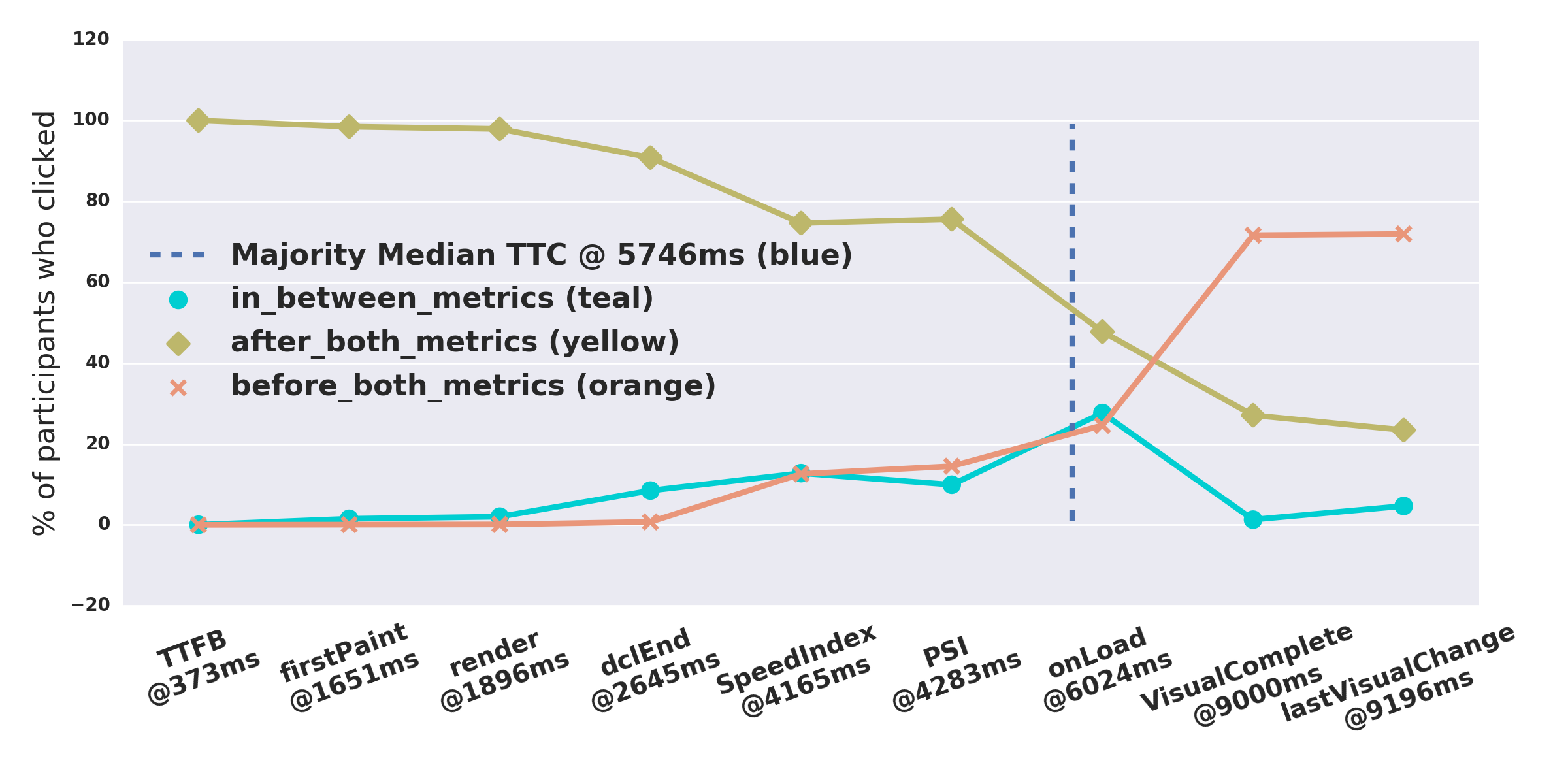}
\vspace{-0.2in}
\caption{\textbf{TTC event as it relates to synthetic metrics. X-axis labels are different timing milestones (evenly spaced for ease of perusal) of a webpage loading. Y-axis is the percentage of users who pressed a button to indicate their choice: before/between/after the timing events associated with video pair.}}
\vspace{-0.2in}
\label{fig:ttc}
\end{figure}

\vspace{0.2in}
\subsection{Time to Click (TTC)}
\emph{Time to Click} (TTC) was measured from the start of the pairwise video display till the time
when user clicks on a button to indicate their choice. 
TTC informs us when user believes they have sufficient information to make a judgment on perceived speed difference between the two webpages 
being shown. 

\textbf{Figure \ref{fig:ttc}} shows the median position of TTC event, among votes that align with majority choice, as it relates to the median of synthetic metrics across the entire dataset. The position of synthetic metrics in this figure is evenly spaced to make visual inspection easier.
The median TTC, for majority votes across all valid sessions, is at 5746ms. From \textbf{Phase-1} dataset~\cite{github}, median TTC is close to median onLoad but not exactly the same. Phase-1 experiment didn't account for the small variability in human visuomotor response across various device types (say smartPhone to tablet). In future work, we plan fix this gap and build better estimates for TTC.

It is worth noticing that \emph{almost} all participants voted \emph{after} the \emph{Render} event of both webpages. 
On the other hand, decision patterns start to shift between \emph{Render} and \emph{visualComplete} events. Most participants waited until post-onLoad of at least one webpage video.
Yet, very few waited until the \emph{visualComplete} event --- thus confirming our hypothesis-3 (Section 3.1).


\subsection{Matching Human Perception}
\begin{figure}
\graphicspath{ {/} }
\centering
\includegraphics[scale=0.3]{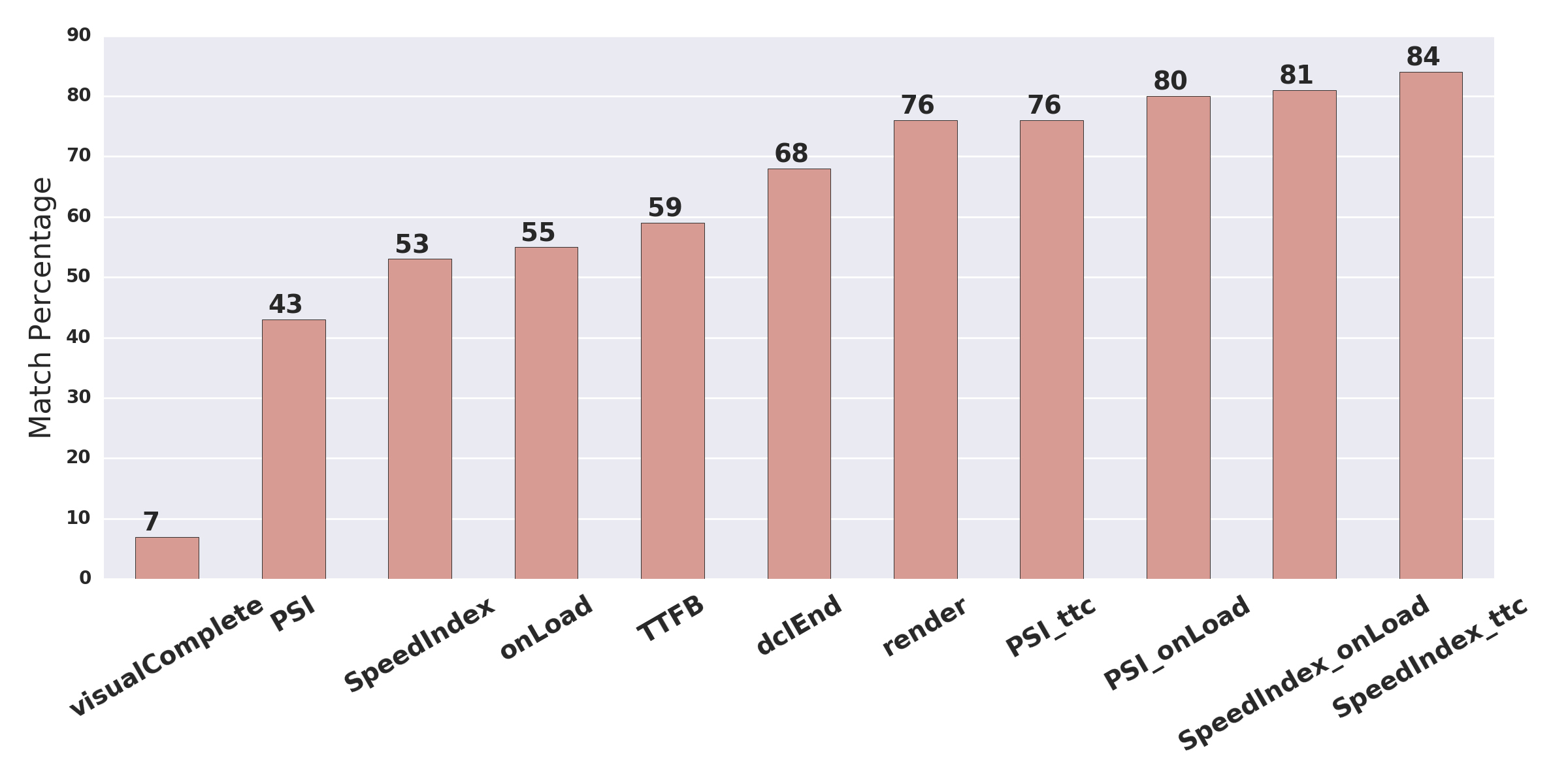}
\vspace{-0.2in}
\caption{\textbf{Percentage match of synthetic metrics matched to end-user votes on perceived speed, in ranked order. SI and PSI using onLoad and TTC as end points are also included to demonstrate the significant improvement.}}
\vspace{-0.2in}
\label{fig:perc_match}
\end{figure}

To determine how well various synthetic metrics explain user perception, we computed the fraction of the ``synthetic votes'' (associated with each metric) that match real user population's majority votes. \textbf{Figure \ref{fig:perc_match}} validates our first hypothesis, that there does not exist a single metric that can explain users perception above $90\%$ accuracy. While ``onLoad'' only matches 55\% of the majority vote, original SI (integrated up to ``visualComplete'') is even worse (53\% match). 

The top 5 metrics that best match user perception on IR-500 webpage speed are 
\emph{$SI_{TTC}$}, \emph{$PSI_{TTC}$}, \emph{$SI_{onLoad}$}, \emph{$PSI_{onLoad}$} (i.e.,\emph{SI} and \emph{PSI} integrated up to ``onLoad'' and ``TTC (majority)'' - \textbf{Equation 2}) and \emph{Render}.
This validates our hypothesis-2 about visual metrics (Section 3.1).

\subsection{Joint ML model}
Although we can explain 84\% of the majority perception using \emph{$SI_{TTC}$}, 
the percentage match score can only serve as an empirical observation. We also didn't want to solely rely on \emph{SI}, knowing that it doesn't account for layout instability that significantly impacts human user QoE. To make our findings more actionable and robust, we tried simple supervised ML modeling approaches using all synthetic metrics to build a predictive model for user perception of speed. 
The predictive label of the model is three options that we provided for our participants. Model features are constructed from the normalized difference of each 
metric. All models were trained and tested using 10-fold cross validation.
We show results from two state-of-art classification models: Random Forest \cite{breiman2001random} and Gradient Boosting \cite{friedman2001greedy}. 

\begin{figure}
\graphicspath{ {/} }
\centering
\includegraphics[scale=0.35]{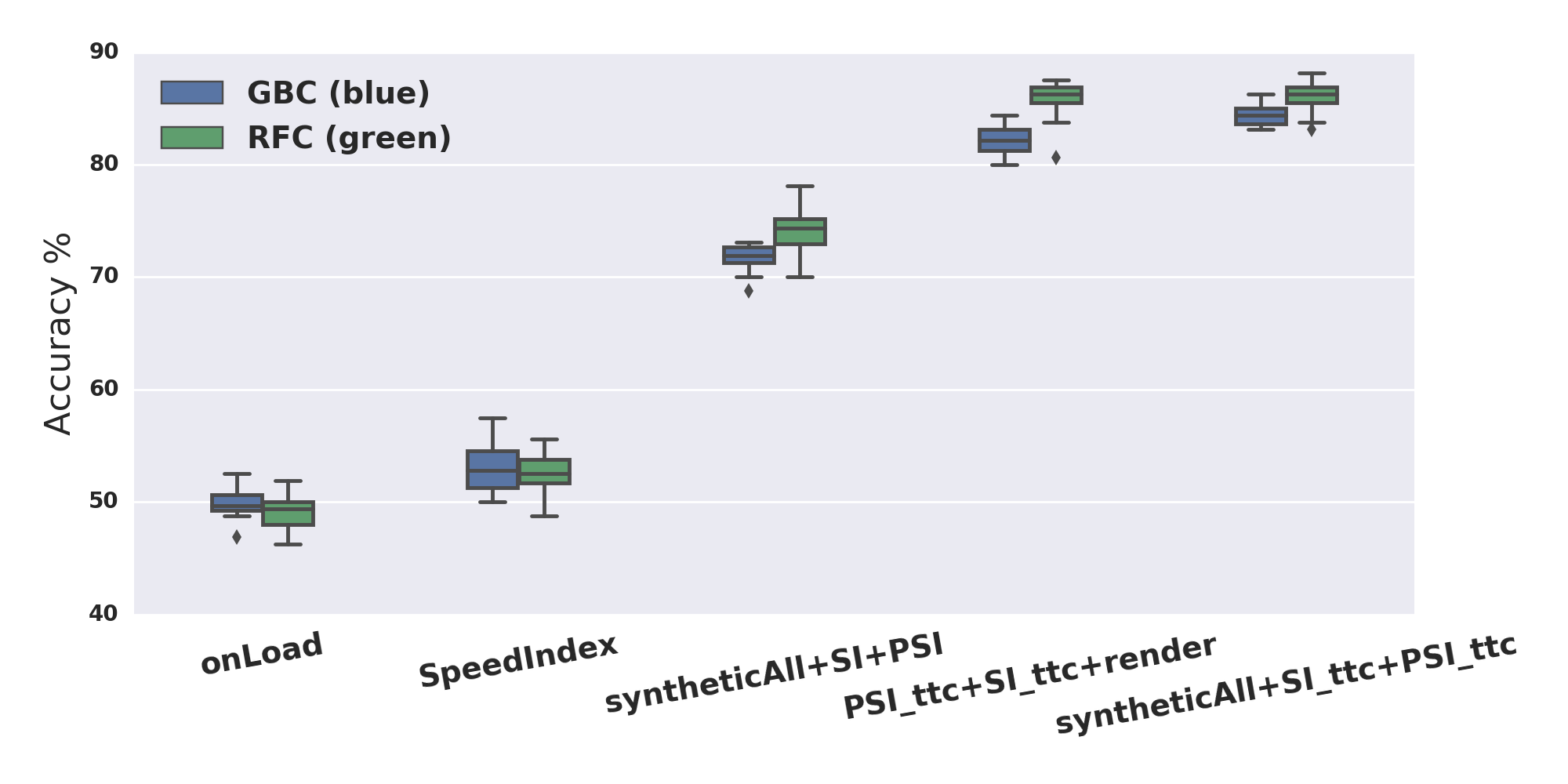}
\vspace{-0.3in}
\caption{\textbf{Box plot of ML models predicting majority vote of human users using different features. All models were trained and tested using 10-fold cross validation.}}
\vspace{-0.2in}
\label{fig:model}
\end{figure}

We can build a very expansive set of models using permutations of different metrics and features. Due to the limited space in this paper, we only show the illustrative results from 
\begin{enumerate*}
\item $onLoad$,
\item $SI$,
\item All synthetic metrics (noted as syntheticAll in the plot) + $SI$ + $PSI$,
\item $PSI_{TTC}$ + $SI_{TTC}$ + $render$,
\item syntheticAll + $PSI_{TTC}$ + $SI_{TTC}$.
\end{enumerate*}
\textbf{Figure \ref{fig:model}} provides us a clear view that a joint model of all synthetic metrics without any fine tuning can predict users' speed-based QoE choices at the accuracy level of 70\% to 75\%, which is an improvement compared to the onLoad or SpeedIndex based models. Despite such significant jump from 50\%+ to 70\%+, the \emph{syntheticAll+SI+PSI} model uses original SI and PSI, which aren't the best (Section 5.2). We then fitted an alternative model replacing SI and PSI with our modified $SI_{TTC}$ and $PSI_{TTC}$.
The new model achieves an accuracy ranges from 87\% to 90\%. In fact, using only three visual metrics ($SpeedIndex_{TTC}$, $PSI_{TTC}$ and $render$) can achieve almost the same level of accuracy as all metrics combined. 

A lot of content relevant to visual QoE is rendered after onLoad --- which explains why joint ML models using onLoad alone do poorly. On the other hand, many websites load visual jitter (such as carousals and pop-ups) after onLoad --- which explains why visual-change aggregation beyond TTC is not that useful for SI or PSI calculations. We speculate that integrating visual change beyond TTC inserts noise into the computation, since a lot of visual jitter happens post-TTC.

\section{Summary and Discussion}
We presented a novel large-scale crowdsourcing framework (\emph{SpeedPerception}) for A/B comparison of end-user QoE. Code for replicating the experimental framework as well as the crowdsourced benchmark data~\cite{github} are freely available, and serve as a useful basis for investigating better QoE metrics in future. 

\emph{SpeedPerception} Phase-1 study on IR 500 websites enabled us to analyze the associations between various web performance metrics and end-user judgments on perceived speed for ATF content. We introduced \emph{Perceptual SpeedIndex} (PSI), and a few systematic variations of both SI and PSI --- all of which serve as key indicators for real user perception of QoE. Phase-1 results showed that while no single performance metric reflected user judgments perfectly, $SI_{TTC}$ and $PSI_{TTC}$ appear to be able to match about 80\% of the majority votes. Moreover, our joint machine learning models predict majority opinions above an accuracy of 85\%. 

In future, we plan to investigate if our findings on human end-user QoE vary from IR 500 ranked websites to Alexa 1000 ranked websites, as well as from desktop webpages to mobile webpages. 

While we showed improved performance using joint ML models, the models we used here don't lend themselves to easy interpretation.
To address this, we plan to develop interpretable models to infer rules that relate webpage structure and page-level metrics to end-user QoE.
\section{Acknowledgments}
We thank Patrick Meenan (WebPagetest.org), Estelle Weyl and Ilya Grigorik for evangelizing \emph{SpeedPerception} and generating significant user participation. We thank Paul Irish, Pierre-Marie Dartus, Shubhie Panicker, and Addy Osmani for independently evaluating and porting PSI for Google Chrome Lighthouse project.
%
\scriptsize{}


\begin{thebibliography}{10}
\bibliographystyle{abbrv}
\bibitem{lighthouse}
{Google Chrome Lighthouse Project} https://github.com/googlechrome/lighthouse.

\bibitem{speedIndex}
{SpeedIndex}
  https://sites.google.com/a/webpagetest.org/docs/using-webpagetest/metrics/speed-index.

\bibitem{github}
{SpeedPerception Benchmark and Results}
  https://github.com/pahammad/speedperception.

\bibitem{catchpoint}
{The Very Real Performance Impact on Revenue}
  http://blog.catchpoint.com/2017/01/06/performance-impact-revenue-real/.

\bibitem{ahammad2015flexible}
P.~Ahammad, R.~Gaunker, B.~Kennedy, M.~Reshadi, K.~Kumar, A.~Pathan, and
  H.~Kolam.
\newblock A flexible platform for qoe-driven delivery of image-rich web
  applications.
\newblock In {\em IEEE ICME 2015}, pages 1--6, 2015.

\bibitem{balachandran2014modeling}
A.~Balachandran, V.~Aggarwal, E.~Halepovic, J.~Pang, S.~Seshan,
  S.~Venkataraman, and H.~Yan.
\newblock Modeling web quality-of-experience on cellular networks.
\newblock In {\em ACM MobiCom 2014}, pages 213--224, 2014.

\bibitem{bocchi2016measuring}
E.~Bocchi, L.~De~Cicco, and D.~Rossi.
\newblock Measuring the quality of experience of web users.
\newblock {\em ACM SIGCOMM Computer Communication Review}, 46(4):8--13, 2016.

\bibitem{breiman2001random}
L.~Breiman.
\newblock Random forests.
\newblock {\em Machine learning}, 45(1):5--32, 2001.

\bibitem{butkiewicz2011understanding}
M.~Butkiewicz, H.~V. Madhyastha, and V.~Sekar.
\newblock Understanding website complexity: measurements, metrics, and
  implications.
\newblock In {\em ACM SIGCOMM IMC}, pages 313--328, 2011.

\bibitem{6smarketing}
E.~Carbery.
\newblock {Website Performance: The Need for Speed}
  http://www.6smarketing.com/blog/website-performance-the-need-for-speed/.

\bibitem{cohen2002prefetching}
E.~Cohen and H.~Kaplan.
\newblock Prefetching the means for document transfer: A new approach for
  reducing web latency.
\newblock {\em Computer Networks}, 39(4):437--455, 2002.

\bibitem{egger2012time}
S.~Egger, P.~Reichl, T.~Ho{\ss}feld, and R.~Schatz.
\newblock ``{T}ime is bandwidth''? {N}arrowing the gap between subjective time
  perception and quality of experience.
\newblock In {\em IEEE ICC 2012}, pages 1325--1330, 2012.

\bibitem{friedman2001greedy}
J.~H. Friedman.
\newblock Greedy function approximation: a gradient boosting machine.
\newblock {\em Annals of statistics}, pages 1189--1232, 2001.

\bibitem{hossfeld2014best}
T.~Hossfeld, C.~Keimel, M.~Hirth, B.~Gardlo, J.~Habigt, K.~Diepold, and
  P.~Tran-Gia.
\newblock Best practices for qoe crowdtesting: Qoe assessment with
  crowdsourcing.
\newblock {\em IEEE Transactions on Multimedia}, 16(2):541--558, 2014.

\bibitem{iyengar2005web}
A.~Iyengar, E.~Nahum, A.~Shaikh, and R.~Tewari.
\newblock Web caching, consistency, and content distribution.
\newblock {\em The Practical Handbook of Internet Computing}, 2005.

\bibitem{kelton17improving}
C.~Kelton, J.~Ryoo, A.~Balasubramanian, and S.~R. Das.
\newblock Improving user perceived page load times using gaze.
\newblock In {\em USENIX NSDI 17}, pages 545--559.

\bibitem{netravali2016polaris}
R.~Netravali, A.~Goyal, J.~Mickens, and H.~Balakrishnan.
\newblock Polaris: Faster page loads using fine-grained dependency tracking.
\newblock In {\em USENIX NSDI 2016}, 2016.

\bibitem{varvello2016eyeorg}
M.~Varvello, J.~Blackburn, D.~Naylor, and K.~Papagiannaki.
\newblock Eyeorg: A platform for crowdsourcing web quality of experience
  measurements.
\newblock In {\em ACM CoNEXT}, pages 399--412, 2016.

\bibitem{wang2016speeding}
X.~S. Wang, A.~Krishnamurthy, and D.~Wetherall.
\newblock Speeding up web page loads with shandian.
\newblock In {\em USENIX NSDI}, 2016.

\bibitem{wang2004image}
Z.~Wang, A.~C. Bovik, H.~R. Sheikh, and E.~P. Simoncelli.
\newblock {\em IEEE transactions on image processing}, 13(4):600--612, 2004.
\end{thebibliography}
%
%
\end{document}